\documentclass[11pt, comsoc]{IEEEtran}

\usepackage{cite}
\usepackage{amsmath,amssymb,amsfonts}
\usepackage{algorithmic}
\usepackage{graphicx}
\usepackage{xcolor}
\usepackage{lipsum}
\usepackage{subcaption}
\usepackage{textcomp}
\usepackage{xcolor}

\begin{document}

\title{Digital Twin Network: Opportunities and Challenges \\
\thanks{This work has been submitted to the IEEE for possible publication. Copyright may be transferred without notice, after which this version may no longer be accessible.}
}

\author{\IEEEauthorblockN{Paul~Almasan\IEEEauthorrefmark{1}, Miquel~Ferriol-Galmés\IEEEauthorrefmark{1}, Jordi~Paillisse\IEEEauthorrefmark{1}, José~Suárez-Varela\IEEEauthorrefmark{1},\\Diego~Perino\IEEEauthorrefmark{2}, Diego~López\IEEEauthorrefmark{2}, Antonio~Agustin~Pastor~Perales\IEEEauthorrefmark{2}, 
Paul~Harvey\IEEEauthorrefmark{3}, Laurent~Ciavaglia\IEEEauthorrefmark{3},\\Leon~Wong\IEEEauthorrefmark{3}, Vishnu Ram\IEEEauthorrefmark{4},
Shihan~Xiao\IEEEauthorrefmark{5},
Xiang~Shi\IEEEauthorrefmark{5}, Xiangle~Cheng\IEEEauthorrefmark{5},\\Albert~Cabellos-Aparicio\IEEEauthorrefmark{1},
Pere~Barlet-Ros\IEEEauthorrefmark{1}}\\
\IEEEauthorblockA{\IEEEauthorrefmark{1}Barcelona Neural Networking Center, 
Universitat Politècnica de Catalunya\\
}
\IEEEauthorblockA{\IEEEauthorrefmark{2}Telefónica Research \IEEEauthorrefmark{3}Rakuten Mobile}
\IEEEauthorblockA{\IEEEauthorrefmark{4}Independent Researcher \IEEEauthorrefmark{5}Huawei Technologies Co.,Ltd.}

}

\maketitle

\begin{abstract}
The proliferation of emergent network applications (e.g., AR/VR, telesurgery, real-time communications) is increasing the difficulty of managing modern communication networks. These applications typically have stringent requirements (e.g., ultra-low deterministic latency), making it more difficult for network operators to manage their network resources efficiently. In this article, we propose the Digital Twin Network (DTN) as a key enabler for efficient network management in modern networks. We describe the general architecture of the DTN and argue that recent trends in Machine Learning (ML) enable  building  a DTN that efficiently and accurately mimics real-world networks. In addition, we explore the main ML technologies that enable  developing the components of the DTN architecture. Finally, we describe the open challenges that the research community has to address in the upcoming years in order to enable the deployment of the DTN in real-world scenarios.
\end{abstract}


\section{Introduction}

In the last years, the digital transformation of both society and industry has led to the emergence of novel network applications. These applications have complex requirements that cannot be easily met by traditional network management solutions, such as network over-provisioning or admission control. For example, novel forms of communication (e.g., AR/VR, holographic telepresence) require ultra-low deterministic latency, while recent industrial developments (e.g., Vehicular Networks) need to adapt to ever-changing network topologies in real-time. At the same time, the number of connected devices is growing rapidly, making modern networks' behaviour highly dynamic and heterogeneous. As a result, modern communication networks have become very complex and costly to manage.

Other industry sectors have recently adopted the Digital Twin (DT) paradigm \cite{8477101} to model complex and dynamic systems. A DT can be understood as a virtual model of a physical object, system, or phenomenon that is represented in the digital world. 
The main advantage of a DT is that it can 
accurately model a complex system without interacting with it, which would otherwise be costly in the physical world. DT examples include enabling smart manufacturing in Industry 4.0 \cite{qi2018digital}, improving the performance and design of complex engineering products (e.g., engine design
) or modelling physical interactions \cite{glatt2021modeling}. 

This article makes the case for the Digital Twin Network (DTN) as a key enabler of efficient control and management of modern real-world networks. Specifically, a DTN allows network operators to design 
network optimization solutions, perform troubleshooting, what-if analysis, or plan network upgrades taking into account the network's expected user growth. 
Since the interaction with the DTN does not require access to the real network, the aforementioned processes can be carried out in real-time, without jeopardizing the physical network.

\begin{figure*}[!t]
  \centering
  \includegraphics[width=0.99\textwidth]{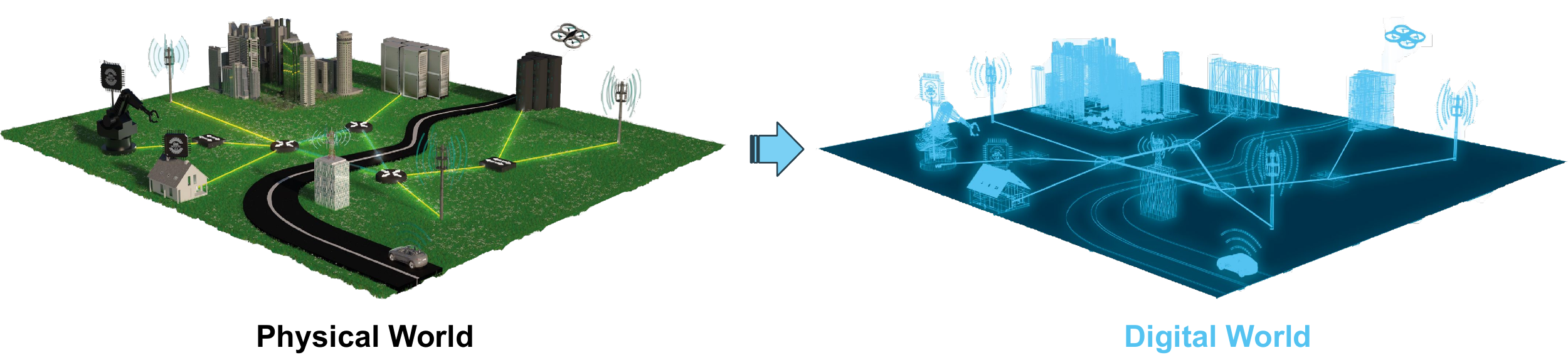}
  \caption{General Overview}
  \label{fig:generalOverview}
\vspace{-0.3cm}
\end{figure*}

In addition, we argue that the latest advances in Machine Learning (ML) make it possible to build a DTN. In particular, recent ML proposals have shown the huge capabilities of modern ML technologies for modeling complex systems. For example,  Alphafold  uses ML to predict protein structures in 3D \cite{senior2020improved}. In computer networks, ML has been successfully used to model path delays \cite{rusek2020routenet}, schedule jobs in data processing clusters \cite{10.1145/3341302.3342080}, and optimize traffic in data-centers \cite{10.1145/3230543.3230551}, among others.

Figure~\ref{fig:generalOverview} shows a general overview of the DTN. First, the digital representation of the physical network
is built in the digital world. Then, the network operator interacts with the DTN in real-time. The DTN is built using already existing ML techniques, making its execution fast and lightweight compared to traditional network simulation technologies (see later). This allows the network operator to immediately obtain accurate performance metrics of the network while evaluating different network configurations on the DTN. 

There is a growing interest in the netwroking community in building a DTN. In particular, Standard Development Organizations (SDO), such as the IETF or the ITU, have started to work on the definition of a DTN \cite{zhou-nmrg-digitaltwin-network-concepts-05}. While their work focuses on defining the main concepts and interfaces of a DTN, in this article we focus on the technologies and research challenges involved in implementing a ML-based DTN, complementing the work of SDOs. 

\section{Applications of the Digital Twin Network}
\subsection{Troubleshooting}
There are many factors that cause network failures (e.g., invalid network configurations, unexpected protocol interactions). Debugging modern networks is complex and time consuming. Currently, troubleshooting is typically done by human experts with years of experience using networking tools.

Network operators can leverage a DTN to reproduce previous network failures, in order to find the source of  service disruptions. Specifically,  network operators can replicate past network failure scenarios and analyze their impact on network performance, making it easier to find specific configuration errors. In addition, the DTN helps in finding more robust network configurations that prevent service disruptions in the future.

\subsection{What-if analysis}
The DTN is a unique tool to perform \textit{what-if analysis}, where the impact of potential scenarios and configurations are safely analyzed using the DTN. In this context, the DTN acts as a \textit{safe sandbox} where different configurations are applied to the DTN to understand their impact on the network. \textit{What-if analysis} helps operators detect network misconfigurations, bottleneck links, observe network performance in case of disasters (e.g., link failures), or predict future network behaviours under different events without any impact on the real network. 

\subsection{Network Planning}
The size and traffic of networks has doubled every year \cite{ellis2016communication}. To accommodate this growth in users and network applications, networks need periodical upgrades. For example, ISPs might be willing to increase certain link capacities or add new connections to alleviate the burden on the existing infrastructure. This is typically a cumbersome process that relies on expert knowledge. Furthermore, modern networks are becoming larger and more complex, thus exacerbating the difficulty of existing solutions to scale to larger networks \cite{10.1145/3452296.3472902}.

The DTN models large infrastructures and produces accurate and fast performance estimates. Hence, it can help in estimating when an existing network will run out of resources, assuming a given growth in users. In addition, its performance estimates are useful to plan the optimal upgrade that can cope with such growth. In short, network operators can leverage the DTN to make better planning decisions and anticipate network upgrades.
 
\subsection{Network Anomaly Detection}
Since the DTN models the behavior of a real-world network, network operators have access to an estimation of the expected network behaviour. When the real-world network behaviour deviates from the DTN's behavior, it can act as an indicator of an anomaly in the real-world network. Such anomalies can appear at different places in a network (e.g., core, edge, IoT), and different data sources can be used to detect such anomalies.

\subsection{Education and Training}
As discussed before, the DTN can be understood as a safe playground where misconfigurations don't affect the real-world system performance. In this context, the DTN can play an important role in improving the education and certification process of network professionals. For example, the DTN can be used in cybersecurity scenarios to evaluate the effects of network attacks and possible counter-measures.

\section{Architecture}

\begin{figure}[!t]
  \centering
  \includegraphics[width=0.99\linewidth]{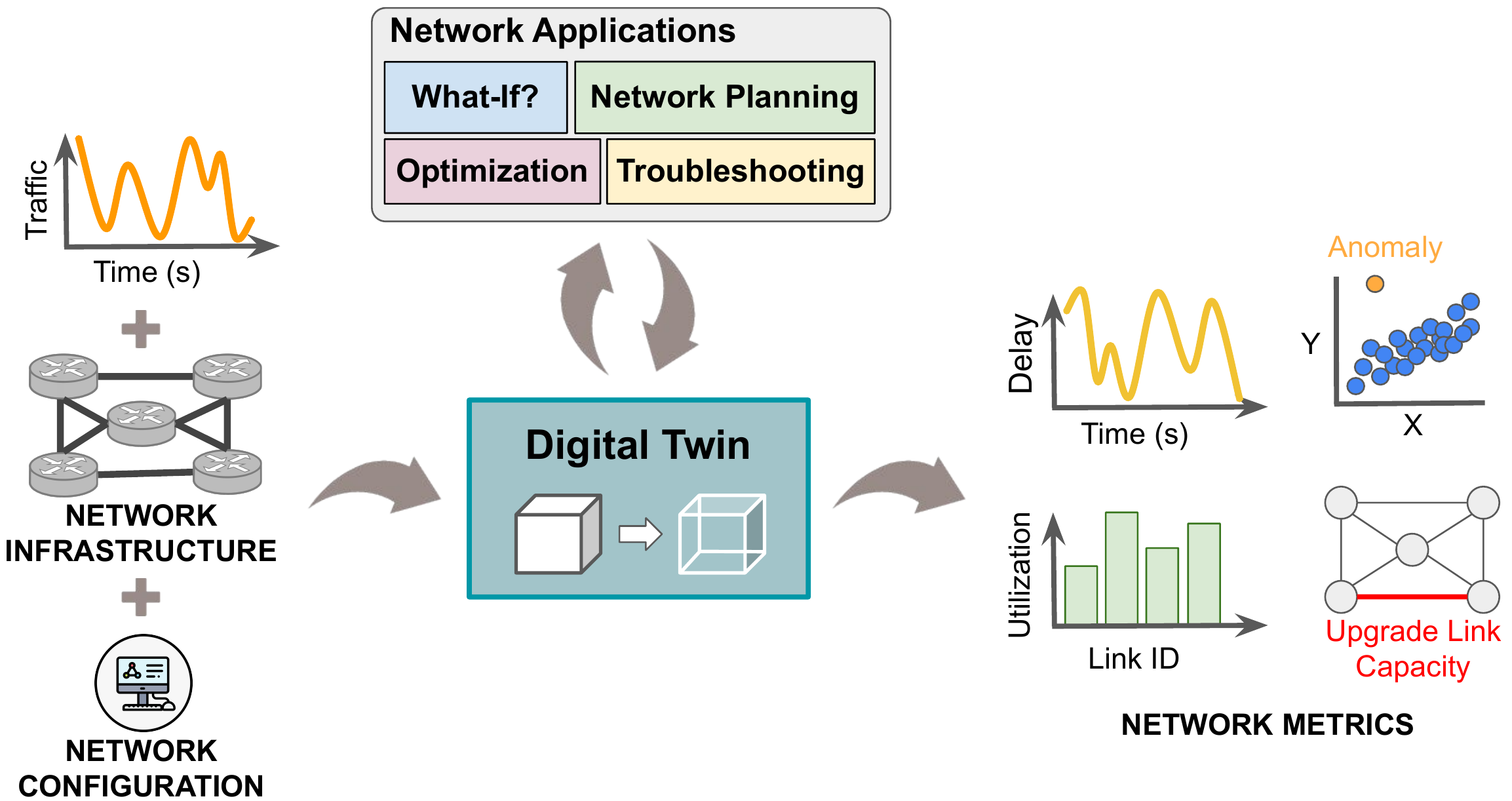}
  \caption{Digital Twin Network Architecture.}
  \label{fig:DTNarchitecture}
\end{figure}

The DTN is a data-driven paradigm that leverages network-related data to build a digital representation of a network. The source of this data is typically from real-world networks, dedicated network testbeds, or network simulation tools. The data might be diverse in nature, but has to  define the network environment or phenomena that the network operator wants to mimic. 

Once the data is gathered, the next step is to build the DTN model. Advances in Deep Learning (DL) technologies showed great modeling capabilities in complex network scenarios \cite{rusek2020routenet,10.1145/3341302.3342080,10.1145/3230543.3230551}. We argue that DL techniques can be leveraged to extract knowledge from the gathered data and build the DTN of a network. 

Figure~\ref{fig:DTNarchitecture} presents the reference architecture of the DTN. The central component of the architecture is the DT, which fundamentally incorporates a network model that mimics the physical network. This model takes as input a set of network-related information and produces a set of performance metrics. In addition, some applications (e.g., Traffic Engineering) can use a network optimizer to find the best solutions to different optimization problems. 

\subsection{Digital Twin Network Model}
The DTN can be seen as a black-box that contains the ML-based network model. This box takes as input some parameters (e.g., traffic, topology, routing, scheduling policies) and outputs some network-related performance metrics (e.g., utilization, delay, packet drops). The inputs and outputs are problem and context dependent. 

The network operator can change the values of the input parameters and observe the resulting performance metrics of the DTN in real-time. This is done without the need of running expensive simulations or instrumenting the real-world network. Since the DTN is a faithful copy of the real-world network, the operator can test any input values, even if these values might cause network service disruptions. This is because the DTN is executed in a safe environment isolated from the real-world network. The output performance metrics of the DTN can be of multiple types (e.g., time-series, per-link prediction, global metric values), which are defined according to the needs of the network operator. 

\begin{figure}[!t]
  \centering
  \includegraphics[width=0.96\linewidth]{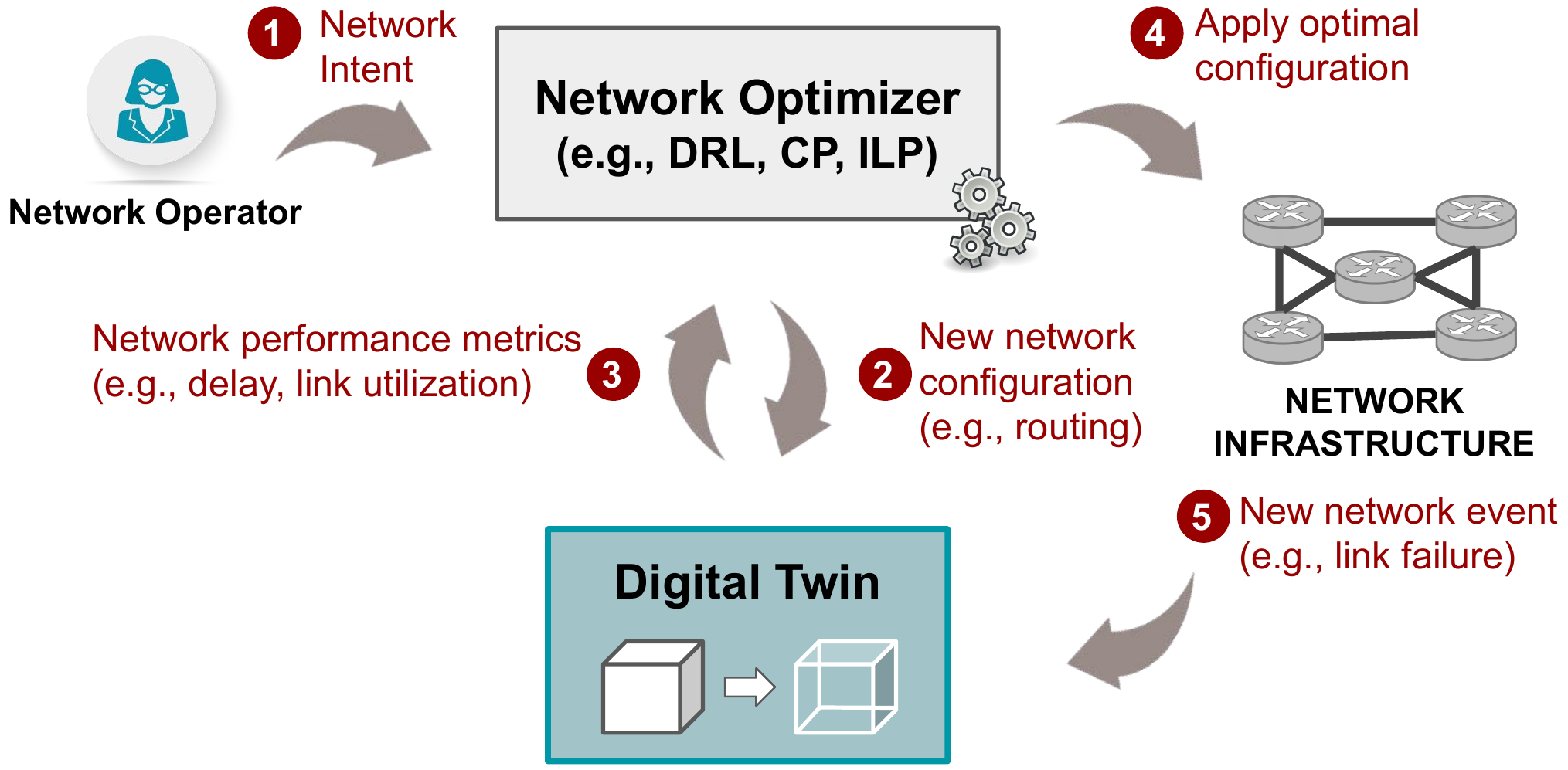}
  \caption{Network optimization process using the Digital Twin Network.}
  \label{fig:optimProcess}
\vspace{-0.1cm}
\end{figure}

\subsection{Network Optimizer}
The DTN is combined with a \textit{network optimizer} to operate a network efficiently.
Figure~\ref{fig:optimProcess} summarizes this process. The optimizer is in charge of searching for the best network configuration (step 2) that fulfills the requirements specified by a network operator (e.g., minimize the maximum link utilization). Such objectives can be expressed using a declarative language. If the performance metrics from the DTN indicate that the solution is not good enough (step 3), the network optimizer continues the search until the stopping condition is met. The best solution found so far is applied directly to the real-world network (step 4). To reduce the DTN's uncertainty, the network configurations can be evaluated using lightweight simulation/emulation tools (e.g., Mininet) before being applied to the real infrastructure. Notice that the optimization process can be closed-loop, where human intervention is not necessarily required.

\subsection{Training the Digital Twin}
Training the DTN model requires building a dataset that contains relevant information of the network. The DTN's accuracy highly depends on the quality of the data, requiring the training dataset to contain a \emph{representative set} of the network behaviour to model. For example, if the goal is to model the delay of network flows, then the dataset has to include a wide range of network scenarios and its impact on the delay. This means including different routing configurations, topologies, scheduling and traffic loads. Likewise, the dataset has to include events that negatively impact the delay, such as link and interface failures, misconfigurations, highly congested scenarios, etc. 

Another important aspect to consider is, \emph{where do we obtain this dataset?} Fundamentally, the dataset can be obtained from real-world networks (i.e., using the real-world network events) or from non-production dedicated network infrastructures (i.e., using pre-defined network events). In other words, the training dataset can be generated at the customer network or at a dedicated testbed not in production. 

However, generating such training sets in production networks is impractical. As we have mentioned, the dataset must contain -among other data- failures, misconfigurations and congested scenarios. Nevertheless, this is not acceptable in a production network, because it  can cause service interruptions. As a result, we envision that the training dataset will be produced in a non-production dedicated network infrastructure, such as a testbed. In the testbed, the network can be configured with different traffic profiles, failures, misconfigurations and errors, as well as a wide range of valid configurations without disrupting users. 

We must note that \emph{online telemetry} should not be confused with generating the training dataset in a testbed. Online telemetry is defined as how information from various data sources is collected using a set of semi-automated processes. Online telemetry from production networks can be used to expand the training dataset. However, the generation of data using not in production testbeds is still required.

The main challenge of generating the dataset using dedicated networks is that the DTN has been trained in a specific testbed, but when deployed it has to operate in a previously unseen customer network. Hence, the DTN has to operate in scenarios that are not specifically included in the training set. As an example, the topology and traffic profile of the customer network might be different from the ones seen during training in the testbed. In the AI domain, the capability of a ML-based model to operate in unseen scenarios is referred to as \emph{generalization}.

\section{Enabling technologies}
\label{sec:enablingTech}

\subsection{Digital Twin Network Model}
Recent works explored the feasibility of applying DL techniques to model computer networks. Earlier works proposed solutions based on traditional Neural Network (NN) architectures (e.g., Multilayer Perceptron, Recurrent NN) for network modeling. Unfortunately, these solutions yielded  poor performance, because classical NN architectures experience difficulties when modeling complex relational relationships (e.g., between routing, topology and end-to-end path delays). In addition, real-world network topologies can change over time due to external factors (e.g., link failures), and traditional ML-based solutions are not able to adapt to such topological (relational) changes.

\begin{figure}[!t]
  \centering
  \includegraphics[width=0.90\linewidth, height=5.2cm]{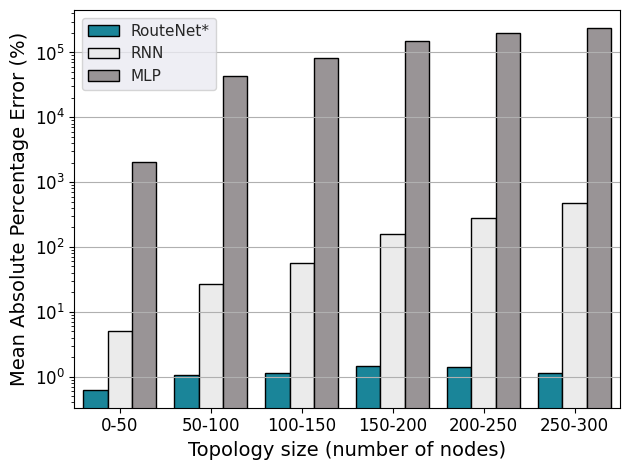}
  \caption{Performance comparison of a GNN based model (RouteNet*), MLP, and RNN for end-to-end path delay prediction for unseen topologies.}
  \label{fig:barplot}
\vspace{-0.5cm}
\end{figure}

Graph Neural Networks (GNN) have been proposed by the ML community to model relational-structured information \cite{4700287}. GNNs are a family of DL models that capture graph dependencies using a message passing algorithm between a graph's entities (e.g., nodes, edges). Since computer networks are fundamentally represented as graphs, GNNs offer  unique advantages for network modeling when compared to traditional NN architectures. 
In the last years, GNNs have demonstrated outstanding performance when  solving network-related problems \cite{rusek2020routenet, 10.1145/3341302.3342080,10.1145/3230543.3230551}. We argue that GNNs are a central technology that enables the construction of scalable network models that can generalize to different network topologies, configurations, and traffic distributions.

Figure~\ref{fig:barplot} shows the results of a motivating experiment to  predict end-to-end path delays using different types of neural networks: a Multi-Layer Perceptron (MLP), a Recurrent Neural Network (RNN), and a GNN (RouteNet*), based on the original RouteNet \cite{rusek2020routenet}. These results demonstrate that GNNs are able to provide accurate estimates even in networks not seen during training. Note that the GNN-based method significantly outperforms the other types of NN architectures. 

\subsection{Network Optimizer}
A network optimizer can leverage a DTN to solve several networking problems (e.g., Traffic Engineering). Specifically, it can use the DTN to obtain immediate network performance estimations during an optimization process. However, the network optimizer must be able to adapt to the real network's dynamics. For example, real-world physical links break due to external factors, or network users can have different  behaviour patterns that cause difficult-to-predict spikes in the utilization of network resources. Thus, it is important to adapt to such changes and to automate the network optimization process in complex scenarios. 

Traditional optimization solutions, such as Constraint Programming (CP) or Integer Linear Programming (ILP), start the optimization process from scratch upon a change in the optimization scenario. This is inefficient because the \textit{knowledge} learned from previous optimization scenarios (e.g., past traffic distribution) is not transferred to the new scenario (e.g., new traffic distribution). Deep Reinforcement Learning (DRL) is a key technology that offers fast operation and can efficiently operate in complex scenarios. DRL differs from traditional optimization solutions because it leverages the knowledge learned in past optimizations. DRL has been applied to network optimization scenarios, showing outstanding performance \cite{10.1145/3452296.3472902, 9651930}. Recent work combining Multi Agent Reinforcement Learning with GNNs showed that the cost of the optimization process scales almost linearly with the size of the network topology \cite{9651930}.

The high complexity of network optimization problems can make the DRL algorithm reach sub-optimal solutions. For example, in Traffic Engineering the number of different routing configurations that minimize a network metric under a threshold (e.g., minimize the maximum link utilization) can be difficult to find. Several works started  combining DRL with traditional optimization methods (e.g., CP) to improve DRL's performance in two different ways. First, traditional optimization methods can be used to teach the DRL agent good actions, helping the agent converge faster to a solution. Second, methods like CP or Local Search can be applied after the DRL optimization process to improve DRL's solution. The latter explores the neighbourhood of the DRL's solution space  to find a better alternative.

\section{Open Research Challenges}

\subsection{Generalization}
The  DTN  should  be  able  to  perform well on different or unseen network scenarios. 
Generalization is important because training a DTN model is not immediate, and it is unfeasible training it after every network change (e.g., link failure). Since these changes  occur very fast (e.g., in vehicular networks), it is not possible to start the training process again. In addition, modern networks are large, making it difficult to replicate them on a testbed, or time consuming to simulate them in order to gather the necessary data to train the DTN model. Therefore, with generalization capabilities it would be possible to train the DTN model in smaller network instances, and deploy it on larger real-world networks without losing significant performance.

\subsection{Flow-based operation}
It is necessary to understand the network traffic at a flow granularity to accurately model a network. However, networks have a large number of flows which raises scalability issues for ML-based methods \cite{10.1145/1879141.1879175}. Some network systems tackle the scalability issue using sampling or aggregation techniques instead of trying to model each individual flow. This enables the network operator to tune the sampling granularity to define how representative the samples are going to be. However, most of the flows are short flows that could be imperceptible to the flow sampling algorithm. Therefore, building flow-based ML models that can operate at flow granularity and at short time scales is a relevant research challenge.

\subsection{Operation in large scenarios}
The efficient operation of large scale networks raises a scalability challenge for the DTN. In such situations, the performance and inference cost should grow at a similar rate as the network size (i.e., larger topologies require larger DTN models, which have higher inference costs). In addition, the training process should also scale well with the network as larger scenarios are inherently more complex, and thus, they require more training time. If a DTN is applied on a larger network and its performance degrades, or it takes more time to execute or to train, the advantages with respect to traditional simulations are lost. 

Operating in large network scenarios implies training a DTN model for large scenarios. However, this is unfeasible as it requires building large and expensive testbeds or executing time-consuming simulations. Recent works that apply ML techniques to networking leverage different techniques to enable scalability. One of them is problem reduction, which reduces the original problem to a smaller instance of the same problem. As an example, graph clustering techniques can be used to reduce the network topology to a smaller version. 

\subsection{Explainability}
As mentioned previously, NNs are typically seen as a black-box. This hinders the deployment of NN-based solutions in real-world network scenarios because network operators struggle to interpret the predictions made by the NN. Recently, the ML community started to investigate how to understand and interpret better DL models. Network operators can leverage the knowledge extracted from the DL trained models to identify potential issues or design better solutions. However, the tools to interpret NN-based systems should be easy to use and friendly for non-ML experts, like network engineers. The networking community is already investigating methods to interpret NNs \cite{10.1145/3387514.3405859}. 

\subsection{Uncertainty estimation}
When a NN is evaluated, it is difficult to obtain a confidence interval of the predicted values. This is important because the NN can produce overconfident predictions about the values or actions (e.g., in DRL), making it difficult for network operators to trust NN-based solutions, and to adopt them in real-world deployments. Such limitation is important as network operators usually seek robust and reliable methods to solve networking problems. To enable the deployment of NNs in real-world applications it is necessary to find a method to assess the confidence of the values predicted by the NN. Recent works are trying to solve this problem by substituting the weights from a NN by distributions \cite{blundell2015weight}.

\subsection{Data collection and storage}
In a networking context, the collection and processing of data is challenging and expensive. Data is only valuable if it is meaningful, which is typically achieved by using a common data format or labelling. However, in real-world networks data comes from different sources and  has different formats. Thus, the data collection process has to aggregate or transform the original data to a common representation, which is decoupled from the data source. This requires using common telemetry systems to gather relevant network-related data.

One of the limitations of gathering network related data is that it can require an extremely large amount of storage. For example, in production scale data-centers (with the order of thousands of servers) the majority of flows are short flows, i.e. flows that have a very short life \cite{10.1145/1879141.1879175}. Storing the data of all flows can take hundreds of GB \cite{10.1145/1879141.1879175}, which can become unfeasible to store and to process. This calls for the research community to find a method to reduce the size of the data and to study network compression techniques.

\section{Conclusion}

In this article we introduced the DTN paradigm. We argued that the DTN enables the design and development of efficient network management tools for modern networks. Recent advances in ML enable building a DTN that is able to mimic the behaviour of a real-world network. We believe that existing ML techniques  allow the networking industry to build market-ready DTNs. While there are still some open challenges to be addressed for a full-scale DTN deployment in real networks, we encourage the networking community to explore solutions to these challenges.

\section*{Acknowledgment}

This publication is part of the Spanish I+D+i project TRAINER-A (ref. PID2020-118011GB-C21), funded by MCIN/AEI/10.13039/501100011033. This work is also partially funded by the Catalan Institution for Research and Advanced Studies (ICREA) and the Secretariat for Universities and Research of the Ministry of Business and Knowledge of the Government of Catalonia and the European Social Fund.

\bibliographystyle{IEEEtran}
\bibliography{references}

\begin{thebibliography}{10}
\providecommand{\url}[1]{#1}
\csname url@samestyle\endcsname
\providecommand{\newblock}{\relax}
\providecommand{\bibinfo}[2]{#2}
\providecommand{\BIBentrySTDinterwordspacing}{\spaceskip=0pt\relax}
\providecommand{\BIBentryALTinterwordstretchfactor}{4}
\providecommand{\BIBentryALTinterwordspacing}{\spaceskip=\fontdimen2\font plus
\BIBentryALTinterwordstretchfactor\fontdimen3\font minus
  \fontdimen4\font\relax}
\providecommand{\BIBforeignlanguage}[2]{{%
\expandafter\ifx\csname l@#1\endcsname\relax
\typeout{** WARNING: IEEEtran.bst: No hyphenation pattern has been}%
\typeout{** loaded for the language `#1'. Using the pattern for}%
\typeout{** the default language instead.}%
\else
\language=\csname l@#1\endcsname
\fi
#2}}
\providecommand{\BIBdecl}{\relax}
\BIBdecl

\bibitem{8477101}
F.~Tao, H.~Zhang, A.~Liu, and A.~Y.~C. Nee, ``Digital twin in industry:
  State-of-the-art,'' \emph{IEEE Trans. on Indust. Inform.}, vol.~15, no.~4,
  pp. 2405--2415, 2019.

\bibitem{qi2018digital}
Q.~Qi and F.~Tao, ``Digital twin and big data towards smart manufacturing and
  industry 4.0: 360 degree comparison,'' \emph{IEEE Access}, vol.~6, pp.
  3585--3593, 2018.

\bibitem{glatt2021modeling}
M.~Glatt, C.~Sinnwell, L.~Yi, S.~Donohoe, B.~Ravani, and J.~C. Aurich,
  ``Modeling and implementation of a digital twin of material flows based on
  physics simulation,'' \emph{Jour. of Manuf. Syst.}, vol.~58, pp. 231--245,
  2021.

\bibitem{senior2020improved}
A.~W. Senior, R.~Evans, J.~Jumper, J.~Kirkpatrick, L.~Sifre, T.~Green, C.~Qin,
  A.~{\v{Z}}{\'\i}dek, A.~W. Nelson, A.~Bridgland \emph{et~al.}, ``Improved
  protein structure prediction using potentials from deep learning,''
  \emph{Nature}, vol. 577, no. 7792, pp. 706--710, 2020.

\bibitem{rusek2020routenet}
K.~Rusek, J.~Su{\'a}rez-Varela, P.~Almasan, P.~Barlet-Ros, and
  A.~Cabellos-Aparicio, ``Routenet: Leveraging graph neural networks for
  network modeling and optimization in sdn,'' \emph{IEEE JSAC}, vol.~38,
  no.~10, pp. 2260--2270, 2020.

\bibitem{10.1145/3341302.3342080}
H.~Mao, M.~Schwarzkopf, S.~B. Venkatakrishnan, Z.~Meng, and M.~Alizadeh,
  ``Learning scheduling algorithms for data processing clusters,'' in
  \emph{Proc. of the ACM Special Interest Group on Data Comm.}, ser. SIGCOMM
  '19, New York, USA, 2019, p. 270–288.

\bibitem{10.1145/3230543.3230551}
L.~Chen, J.~Lingys, K.~Chen, and F.~Liu, ``Auto: Scaling deep reinforcement
  learning for datacenter-scale automatic traffic optimization,'' in
  \emph{Proc. of the ACM Special Interest Group on Data Comm.}, ser. SIGCOMM
  '18, New York, USA, 2018, p. 191–205.

\bibitem{zhou-nmrg-digitaltwin-network-concepts-05}
\BIBentryALTinterwordspacing
C.~Zhou, H.~Yang, X.~Duan, D.~Lopez, A.~Pastor, Q.~Wu, M.~Boucadair, and
  C.~Jacquenet, ``{Digital Twin Network: Concepts and Reference
  Architecture},'' IETF, Internet-Draft
  draft-zhou-nmrg-digitaltwin-network-concepts-05, Oct. 2021, (Accessed on Nov.
  22, 2021). [Online]. Available:
  \url{https://datatracker.ietf.org/doc/html/draft-zhou-nmrg-digitaltwin-network-concepts-05}
\BIBentrySTDinterwordspacing

\bibitem{ellis2016communication}
A.~Ellis, N.~M. Suibhne, D.~Saad, and D.~Payne, ``Communication networks beyond
  the capacity crunch,'' \emph{Phil. Trans. R. Soc. A}, 2016.

\bibitem{10.1145/3452296.3472902}
H.~Zhu, V.~Gupta, S.~S. Ahuja, Y.~Tian, Y.~Zhang, and X.~Jin, ``Network
  planning with deep reinforcement learning,'' in \emph{Proc. of the ACM
  SIGCOMM Conf.}, ser. SIGCOMM '21.\hskip 1em plus 0.5em minus 0.4em\relax New
  York, USA: ACM, 2021, p. 258–271.

\bibitem{4700287}
F.~Scarselli, M.~Gori, A.~C. Tsoi, M.~Hagenbuchner, and G.~Monfardini, ``The
  graph neural network model,'' \emph{IEEE Trans. on Neural Netw.}, vol.~20,
  no.~1, pp. 61--80, 2009.

\bibitem{9651930}
G.~Bernárdez, J.~Suárez-Varela, A.~López, B.~Wu, S.~Xiao, X.~Cheng,
  P.~Barlet-Ros, and A.~Cabellos-Aparicio, ``Is machine learning ready for
  traffic engineering optimization?'' in \emph{2021 IEEE 29th International
  Conference on Network Protocols (ICNP)}, 2021, pp. 1--11.

\bibitem{10.1145/1879141.1879175}
T.~Benson, A.~Akella, and D.~A. Maltz, ``Network traffic characteristics of
  data centers in the wild,'' in \emph{Proc. of the ACM SIGCOMM Conf. on
  Internet Measur.}, ser. IMC '10, New York, USA, 2010, p. 267–280.

\bibitem{10.1145/3387514.3405859}
Z.~Meng, M.~Wang, J.~Bai, M.~Xu, H.~Mao, and H.~Hu, ``Interpreting deep
  learning-based networking systems,'' in \emph{Proc. of the ACM Special
  Interest Group on Data Comm. on the App., Tech., Arch. and Prot. for Computer
  Comm.}, ser. SIGCOMM '20, New York, USA, 2020, p. 154–171.

\bibitem{blundell2015weight}
C.~Blundell, J.~Cornebise, K.~Kavukcuoglu, and D.~Wierstra, ``Weight
  uncertainty in neural network,'' in \emph{Proc. of the 32nd Int'l. Conf. on
  Machine Learning}.\hskip 1em plus 0.5em minus 0.4em\relax PMLR, 2015, pp.
  1613--1622.

\end{thebibliography}

\begin{IEEEbiographynophoto}{Paul Almasan}
is a PhD candidate at the Barcelona Neural Networking Center, Universitat Politècnica de Catalunya.
\end{IEEEbiographynophoto}

\begin{IEEEbiographynophoto}{Miquel Ferriol-Galmés}
is a PhD candidate at the Barcelona Neural Networking Center, Universitat Politècnica de Catalunya.
\end{IEEEbiographynophoto}

\begin{IEEEbiographynophoto}{Jordi Paillisse}
is a postdoctoral researcher at the Barcelona Neural Networking Center, Universitat Politècnica de Catalunya.
\end{IEEEbiographynophoto}

\begin{IEEEbiographynophoto}{José Suárez-Varela}
is a postdoctoral researcher at the Barcelona Neural Networking Center, Universitat Politècnica de Catalunya.
\end{IEEEbiographynophoto}

\begin{IEEEbiographynophoto}{Diego Perino}
is the Director of Telefónica Research, a team of researchers and technical experts in the areas of Artificial Intelligence, Networks and Systems, Security and Privacy and Human computer Interaction.
\end{IEEEbiographynophoto}

\begin{IEEEbiographynophoto}{Diego López}
is a senior technology expert at Telefonica I+D, Chair of ETSI NFV and PDL ISGs.
\end{IEEEbiographynophoto}

\begin{IEEEbiographynophoto}{Antonio Agustin Pastor Perales}
works as an expert in Telefónica I+D for global technical areas where he  is involved in network and security innovation activities.
\end{IEEEbiographynophoto}

\begin{IEEEbiographynophoto}{Paul Harvey}
is the research lead and co-founder of the Rakuten Mobile Innovation Studio.
\end{IEEEbiographynophoto}

\begin{IEEEbiographynophoto}{Laurent Ciavaglia}
is a researcher at Rakuten Mobile and Co-Chair of the Network Management Research Group at the Internet Engineering Task Force.
\end{IEEEbiographynophoto}

\begin{IEEEbiographynophoto}{Leon Wong}
is the Industry Research Collaboration Lead and Research Engineering Lead for Research \& Innovation Lab in Rakuten Mobile.
\end{IEEEbiographynophoto}

\begin{IEEEbiographynophoto}{Vishnu Ram}
is an independent researcher with more than two decades in the telecom industry.
\end{IEEEbiographynophoto}

\begin{IEEEbiographynophoto}{Shihan Xiao}
received his PhD degree at Tsinghua University in 2017 and is currently a technical expert of Network AI at Huawei Technologies.
\end{IEEEbiographynophoto}

\begin{IEEEbiographynophoto}{Xiang Shi}
is a senior engineer at the Network Technology Laboratory of Huawei Technologies.
\end{IEEEbiographynophoto}

\begin{IEEEbiographynophoto}{Xiangle Cheng}
is a research fellow working on NetAI technologies at Huawei.
\end{IEEEbiographynophoto}

\begin{IEEEbiographynophoto}{Albert Cabellos-Aparicio}
is a professor at Universitat Politècnica de Catalunya and director at the Barcelona Neural Networking Center.
\end{IEEEbiographynophoto}

\begin{IEEEbiographynophoto}{Pere Barlet-Ros}
is a professor at Universitat Politècnica de Catalunya and scientific director at the Barcelona Neural Networking Center.
\end{IEEEbiographynophoto}

\end{document}